# Self-Driven Graphene Tearing and Peeling: A Fully Atomistic Molecular Dynamics Investigation

Alexandre F. Fonseca[1] and Douglas S. Galvão[1,2]

[1]Applied Physics Department, Institute of Physics "Gleb Wataghin", University of Campinas - UNICAMP, Campinas, São Paulo, CEP 13083-859, Brazil.

[2]Center for Computational Engineering and Sciences, UNICAMP, Campinas, São Paulo, Brazil.

ABSTRACT

*In spite of years of intense research, graphene continues to produce surprising results. Recently, it was experimentally observed that under certain conditions graphene can self-drive its tearing and peeling from substrates. This process can generate long, micrometer sized, folded nanoribbons without the action of any external forces. Also, during this cracking-like propagation process, the width of the graphene folded ribbon continuously decreases and the process only stops when the width reaches about few hundreds nanometers in size. It is believed that interplay between the strain energy of folded regions, breaking of carbon-carbon covalent bonds, and adhesion of graphene-graphene and graphene-substrate are the most fundamental features of this process, although the detailed mechanisms at atomic scale remain unclear. In order to gain further insights on these processes we carried out fully atomistic reactive molecular dynamics simulations using the AIREBO potential as available in the LAMMPS computational package. Although the reported tearing/peeling experimental observations were only to micrometer sized structures, our results showed that they could also occur at nanometer scale. Our preliminary results suggest that the graphene tearing/peeling process originates from thermal energy fluctuations that results in broken bonds, followed by strain release that creates a local elastic wave that can either reinforce the process, similar to a whip cracking propagation, or undermine it by producing carbon dangling bonds that evolve to the formation of bonds between the two layers of graphene. As the process continues in time and the folded graphene decreases in width, the carbon-carbon bonds at the ribbon edge and interlayer bonds get less stressed, thermal fluctuations become unable to break them and the process stops.*

## INTRODUCTION

Graphene became a paradigm of the ideal two-dimensional material. Owning one of the most desirable combination of physical, chemical and structural properties [1-3], graphene and its derivatives (for example, graphene-oxide [4], graphane [5] and graphyne [6]) are considered the ultimate nanostructures for diverse types of applications [7-9].

The special mechanical and structural properties of graphene have been explored in studies involving graphene-metal interfaces [10], strain engineering of its electronic structure [11], tailoring graphene coefficient of thermal expansion [12], mechanical properties of nanocomposites [13], among others [7-9]. In particular, the strength and fracture properties of graphene are key to understand the mechanical limits to which it is possible to take graphene-derived nanodevices [8,14].

Recently, an experiment showed that graphene, under certain conditions, can self-drive its tearing and peeling off from substrates, thus forming long, micrometer size, graphene nanoribbons [15]. In other words, this process can generate long folded nanoribbons without the action of any external forces. One of the conditions for this process to occur is the formation of a hole in a suspended part of graphene, then followed by the creation of an initial fold on the sides of the hole. Thus, the process continues by itself forming final structures similar to flowers, where each petal is a micrometer size graphene nanoribbon that grown by itself through the above process. The number of petals depends on the form of the hole. The final size of the ribbons depends on the width of the graphene folded ribbon that continuously decreases until reaching about few hundreds nanometers in size.

An interplay between the strain energy of folded regions, breaking of carbon-carbon covalent bonds, and adhesion of graphene-graphene and graphene-substrate were proposed to be the most fundamental features of this process (see, for example, Ref. [16]). However, the detailed mechanisms of this phenomenon at atomic scale remain unclear. In order to gain further insights on these processes, we carried out fully atomistic reactive molecular dynamics simulations using the AIREBO [17] potential as available in the LAMMPS [18] computational package.

In particular, we have addressed the folowing questions: *what is the onset of crack growth and propagation? What is the local atomic structure of broken ribbons? What is the minimum size of ribbon width, at least, to start the self-tearing and self-peeling off process? What is the initial ribbon propagation velocity?* Our results show that the process of self-tearing and self-peeling off graphene nanoribbons can occur at nanometer scale; it is enhanced by thermal energy fluctuations and that the strain release due to tearing can, at the same time, either reinforce the process, similarly to a whip cracking propagation [19], or undermine it by producing carbon dangling bonds that form new bonds between the two layers of graphene, thus stoping the ribbon propagation.

**THEORY AND SIMULATION DETAILS**

Classical molecular dynamics simulations of the structures shown in Figure 1 were performed with the AIREBO potential [17], as available in LAMMPS code [18]. Structures having zigzag and armchair crack edges were considered.

The structures are first geometry optimized by energy minimization methods (with force tolerance of $10^{-8}$ eV/Å) following by MD simulations at 300 K and 600 K for several nanoseconds using a Langevin thermostat. Some structures were also simulated at 1000 K in order to verify stability and thermal fluctuation effects. Different width sizes, *w*, were considered. Extremities were kept fixed to mimic large sized graphene structures.

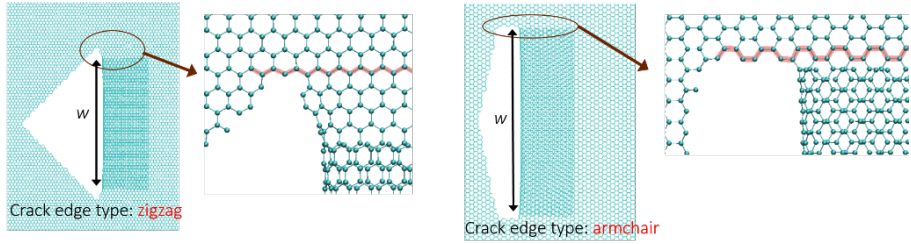

**Figure 1.** Structural models of a folded graphene nanoribbon cut along two types of edges, zigzag (left panel) and armchair (right panel), considered in this study. $w$ is the initial ribbon width. The magnification of the local crack edge structures inside the circles are presented on the right of each panel. Pink lines are shown to indicate the chirality of the edge.

Substrates with different energies were simulated in order to test the results against attractive forces. Substrate energy definition, $E$, is given below, with the parameters $\sigma$ and $r_c$ fixed in 3.5 Å and 12 Å, respectively. Simulations were performed for several values of the parameter $\varepsilon$.

$$E = 4\varepsilon \left[ (\sigma/r)^{12} - (\sigma/r)^{6} \right] \quad r < r_c . \qquad (1)$$

## RESULTS AND DISCUSSION

The first set of results concern the structures whose ribbon widths are not enough to start to the self-tearing and self-peeling off process, even being simulated at high temperatures. Some snapshots of these structures simulated at 600 K are shown in Figure 2.

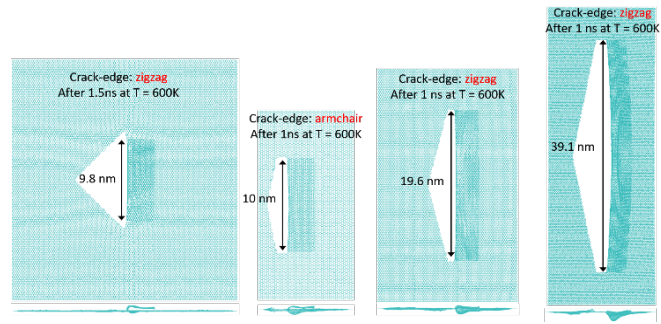

**Figure 2.** Upper and lateral views of the snapshots of the structures with width sizes that are not enough to promote self-tearing and self-peeling off of graphene nanoribbon. The type of chirality of the crack-edge and the simulation details regarding the snapshot are shown in the figure.

When considering a system with ribbon width $w = 80$ nm, the process of self-tearing and self-peeling off was observed. In Figure 3 below, we present the results for the structure with the zigzag crack-edge. Similar results were obtained for the armchair crack-edge.

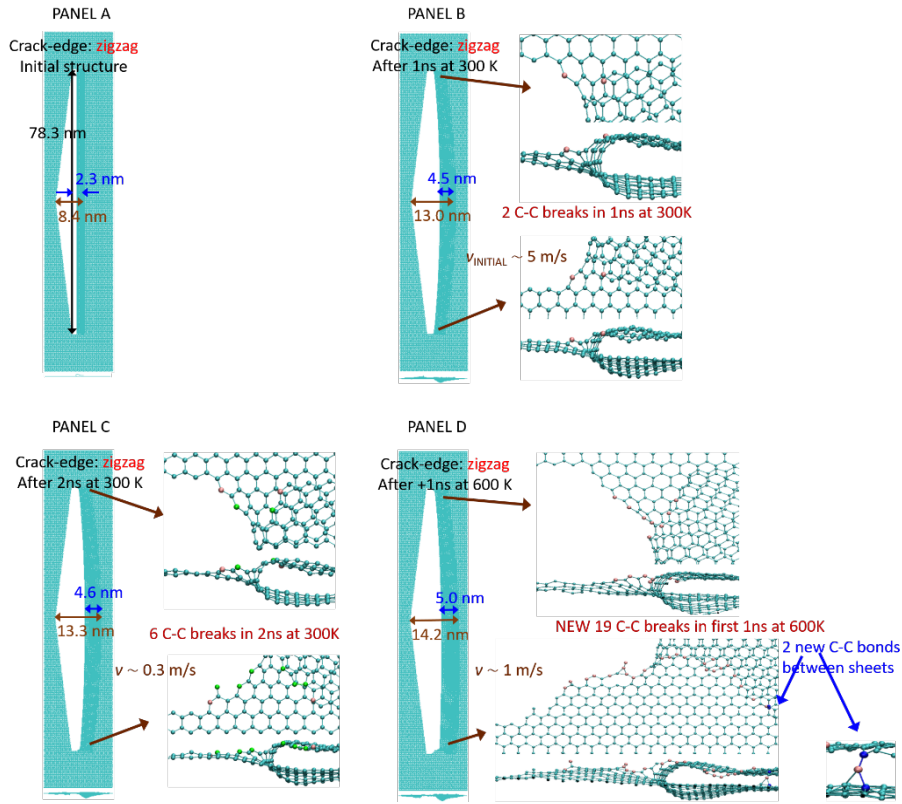

**Figure 3**. Atomic structural details (upper and lateral views) of the smallest width size structure for which crack propagation was observed. Besides the initial optimized structure (PANEL A), structures are shown at every 1 ns of simulation at 300 K (PANELS B and C), and at 600 K (PANEL D). Magnifications of the fractured part of the structures are shown on their right. Red and green atoms are carbon atoms with broken bonds from the crack propagation. Blue atoms in PANEL D are those that formed bonds between the two graphene sheets. The estimated velocity of the front of the graphene nanoribbon and the computed number of broken carbon-carbon bonds at the edges are also shown. Similar results were obtained for structures with armchair crack-edges.

Panel A of Figure 3 shows the initial structure as just geometry optimized. It has 78.3 nm of width, 2.3 nm of folded nanoribbon layer and 8.4 nm of distance between the ribbon edge and the opposite hole edge. These two distances are measured in order to estimate the ribbon propagation velocity. Panels B and C show the MD snapshots of the structure simulated after 1 ns and 2 ns, respectively, at 300 K. During the first 1 ns of simulation at 300 K, just 2 carbon-carbon (C-C) bonds are broken, one at each side of the ribbon lateral edge (Panel B of Figure 3). In the magnifications of the lateral edges, red carbon atoms are those which broken bonds due to the self-tearing and self-peeling off processes during this first 1ns of simulation. The front edge of the ribbon moved by about 4.6 nm, what gives a rough estimate of 5 nm/ns or 5 m/s. This large value comes not from the breaking/tearing of the C-C bonds, but from the adjustment of the stacking of upper and bottom layers of graphene. Panel C shows additional 1 ns of simulation at 300 K. During this time, new 6 C-C bonds are broken and the front of the ribbon edge moved away only 0.3 nm and the ribbon propagation velocity significantly decreases.

An interesting behaviour was observed when the temperature increased from 300 to 600 K (Panel D of figure 3). Many more C-C bonds are broken after 1 ns, the propagation velocity of the front of the ribbon increased again, but an unexpected C-C bond formation was observed between the graphene layers. Two C-C bonds formed between the layers are what prevented further further ribbon propagation.

The effects of thermal fluctuations can be estimated by comparing the structure and ribbon propagation velocities from Panels A to B and from Panels C to D. They allow the system to achieve smaller energy stacking configurations by inducing relative layer movements. Also, the thermal fluctuations increase the rate of bond breaking as seen in Panel D. However, large thermal fluctuations can also lead to the formation of new C-C bonds between the graphene layers, thus preventing further ribbon propagation.

An important observation is that formed line of broken bonds results in the decrease of the ribbon width, what is consistent with the experimental results.

Another point observed in figure 3 is that the tearing occurs neither uniformly nor symmetrically. The magnification of the upper lateral crack edge shown in Panel D shows a zigzag pattern of the edge formed by the broken bonds. But the magnification of the bottom lateral crack edge shows a mixture of zigzag, armchair and some dangling carbon atoms at the edge formed by broken bonds.

The effects of the substrates were also considered. For $\varepsilon = 0.01$ eV (see equation (1)), the same MD simulations for the same initial structure shown in Panel A of Figure 3, provided the following results after the first 1 ns at 600 K: decrease of the front ribbon growth velocity (from 1 m/s to 0.48 m/s); and decrease of the carbon-carbon broken bonds (none at 300 K, and only 7 after first 1 ns at 600 K). So, the effect of the substrate is only to decrease the rate of the ribbon propagation as induced by self-tearing and self-peeling off processes.

**CONCLUSION**

In this study we present the first MD simulation of the process of self-driven tearing and peeling off graphene nanoribbons. We showed that folded ribbons having about 80 nm are sufficient to start the process and that thermal fluctuations play important rule on the stacking configuration and in the resulting number of broken carbon-carbon bonds. Thermal fluctuations also were shown to lead to undesirable formation of carbon-carbon bonds between the graphene layers, which prevents further ribbon propagation. These results suggest that as long as the folded graphene decreases in width, the carbon-carbon bonds at the ribbon edge, as well as the interlayer bonds get less stressed, thermal fluctuations become unable to break these bonds and the tearing and peeling off processes stop.

**ACKNOWLEDGMENTS**

AFF and DSG are fellows of the Brazilian Agency CNPq. AFF acknowledges the grant #2016/00023-9 from São Paulo Research Foundation (FAPESP) and from FAEPEX/UNICAMP. DSG acknowledges the Center for Computational Engineering and Sciences at Unicamp for financial support through the FAPESP/CEPID Grant #2013/08293-7. This research also used the computing resources and assistance of the